\documentclass[%
 reprint,
 amsmath,amssymb,
 aps,
]{revtex4-2}

\usepackage{graphicx}
\usepackage{dcolumn}
\usepackage{enumerate}
\usepackage{physics}
\usepackage{mathtools}
\usepackage{bm}
\usepackage{hyperref}
\usepackage{comment}
\usepackage[dvipsnames]{xcolor}
\usepackage{soul}
\definecolor{g1}{rgb}{0.67, 0.88, 0.69}

\colorlet{h1}{Orange!30}

\begin{document}

\preprint{APS/123-QED}

\title{One-Way Quantum Secure Direct Communication\\with Choice of Measurement Basis as the Secret}

\author{Santiago Bustamante}
\email{santiago.bustamanteq@udea.edu.co}
\affiliation{Instituto de Física, Universidad de Antioquia, Calle 70 No. 52-21, Medellín, Colombia.}

\author{Boris A. Rodríguez}
\affiliation{Instituto de Física, Universidad de Antioquia, Calle 70 No. 52-21, Medellín, Colombia.}

\author{Elizabeth Agudelo}
\affiliation{TU Wien, Atominstitut \& VCQ,  Stadionallee 2, 1020 Vienna, Austria}


\begin{abstract}
Motivated by the question of the distinguishability of ensembles described by the same compressed density operator, we propose a model for one-way quantum secure direct communication using finite ensembles of shared EPR pairs per bit and a public authenticated classical channel, where the local choice of one of two mutually-unbiased measurement bases is the secret bit. 
In this model, both the encoding and decoding of classical information in quantum systems are implemented by measurements in either the computational or the Hadamard basis. 
Using the quantum wiretap channel theory, we study the secure net bit rates and certify information-theoretic security of different {instances} of {the model} when the quantum channel is subject to BB84-symmetric attacks. 
Since no local unitary operations need to be performed by the receiver, the proposed model is suitable for real-life implementations of secure direct communication in star network configurations.
\end{abstract}

\maketitle


\section{Introduction}
\label{sec:intro}

Quantum information science has become a fundamental pillar for future technological advances in areas such as communications~\cite{Zhang2024}, computing~\cite{Sood2024} and cryptography~\cite{Pirandola2020}. 
This is because, at its core, the laws of quantum mechanics offer a framework that challenges and refines classical theories of information processing.
This paradigm shift provides new approaches to numerous technological challenges. 
Particularly, quantum cryptography has emerged as a robust tool for ensuring secure communication between two users who share a quantum channel, and it was the first task in quantum information theory that came to be considered a mature technology, ready for commercialization for over a decade~\cite{Scarani2009}.
Quantum cryptographic protocols, which are mostly based on quantum key distribution, have been shown to offer security guarantees that go beyond those of conventional classical cryptography~\cite{Gisin2002,Shor2000,Wolf2021,Andriolo2026}.

A less explored yet promising paradigm within the field is quantum secure direct communication (QSDC), which seeks to enable secure communication without the need for a shared secret key~\cite{Long2007,Bostrom2002}. 
Many different QSDC protocols have been proposed in the last couple of decades, yet further development, particularly on the tools for information-theoretic security of these protocols, is still required~\cite{Pan2024}. 
Among the different QSDC protocols is the one proposed by Wang Chuan \textit{et al.}, which uses finite ensembles with the same compressed density operator as resources to securely transmit single bits of information~\cite{Wang2006}.
We will from now on refer to this protocol as the CDM06 protocol for convenience (CDM standing for compressed density matrix, in concordance with the authors' convention). 
Although the authors state in their article that their protocol is ``more of conceptual than practical interest" due to its low resourcefulness, in this article we revisit their protocol and expand upon their work, motivated by the foundational question of whether different ensembles described by the same statistical mixture can be distinguished~\cite{Long2006}, and the interesting quantum encoding scheme where the local choice of measurement basis plays the role of the secret bit. Information-theoretic security of the SARG04 quantum key distribution protocol, which uses this same quantum encoding scheme, has been proven for noisy quantum channels~\cite{Sarg04,Cyril2005}, while for QSDC it has been proven only in the zero-error scenario~\cite{Liu2014}. 
This leads us to propose a general model for one-way QSDC where the local choice of measurement basis is the secret bit and prove its information-theoretic security, even in the presence of noise, in accordance to the quantum wiretap channel theory~\cite{Cai2004}.

The article is organized as follows. In Sec.~\ref{sec:distinguishability_problem} we study the underlying secret sharing mechanism of the CDM06 protocol and summarize the discussion on the distinguishability of ensembles with identical compressed operators. 
In Sec.~\ref{sec:wang_protocol} and Sec.~\ref{sec:resourcefulness} we respectively review the CDM06 protocol and identify key characteristics upon which its resourcefulness can be improved. 
In Sec.~\ref{sec:theoretical_model} we propose a general model for one-way QSDC through a state-based mathematical description, and in Sec.~\ref{sec:security} we prove and discuss information-theoretic security of four different {instances} of the model. Finally, conclusions are given in Sec.~\ref{sec:conclusions}.
\section{Framework}

\subsection{Distinguishability of Ensembles with the same Compressed Density Operator} \label{sec:distinguishability_problem}

Before discussing the CDM06 protocol, let us briefly review the evolution of finite ensembles of EPR pairs shared by two parties after a local measurement in the computational or the Hadamard basis, as this underlies the secret sharing mechanism in their protocol.

By \textit{finite ensemble} we refer to a finite collection of $N$ non-interacting identical (yet independently-accessible) quantum systems with an associated Hilbert space $\mathcal{H}$. 
The \textit{compressed density operator} $\varrho$ of a finite ensemble is defined as the state of an average entity of the ensemble, e.g. if there are $N_1,N_2,\dots,N_m$ entities of the ensemble {in the} pure states $\ket*{\psi_1},\ket*{\psi_2},\dots,\ket*{\psi_m}\in\mathcal{H}$ respectively, then $\varrho \propto \sum_{j} N_j \ketbra*{\psi_j}$ is their compressed density operator. 
Notice that $\varrho$ does not generally contain all the information of the ensemble. Given that in our definition of a finite ensemble we assumed that each of its constituents can be accessed independently (at least in principle), then only a \textit{full density operator} $\rho$ acting on $\mathcal{H}^{\otimes N}$ may be able to faithfully contain all information about the system's state~\cite{Long2006}. 
In this sense, differently prepared physical ensembles may share the same compressed density operator; it is only when no correlation among different entities of the ensemble is present that a one-to-one correspondence exists~\cite{ZhanP14,Bustamante2025}.

Finite ensembles of Einstein-Podolsky-Rosen (EPR) pairs {(that is, maximally entangled pairs of qubits}~\cite{Nielsen2010}) represent the basic resource for the transmission of a single bit in the CDM06 protocol, while measurements in the computational and the Hadamard bases implement both the encoding and decoding of classical information in the quantum systems. 
Let us suppose that each of the $N$ EPR pairs {is} shared by two parties, Alice and Bob, in the state
\begin{equation*}
    \ket{\phi_z^+} = \frac{1}{\sqrt{2}} \left ( \ket{\uparrow\uparrow} + \ket{\downarrow\downarrow} \right ) = \frac{1}{\sqrt{2}} \left ( \ket{\rightarrow\rightarrow} + \ket{\leftarrow \leftarrow} \right ),
\end{equation*}
where $\ket{\uparrow}$ ($\ket{\rightarrow}$) and $\ket{\downarrow}$ ($\ket{\leftarrow}$) are the eigenstates of the Pauli $Z$ ($X$) operator with eigenvalues $+1$ and $-1$ respectively. 
Suppose now that Alice measures all of her qubits in the computational ($Z$) basis, {which collapse into the random state} $\ket{k_0k_1 \dots k_{N-1}}$ where each $k_j$ for $j=0,\dots,N-1$ is independently either $\uparrow$ or $\downarrow$ with equal probability $1/2$. 
Due to the quantum correlations present in each EPR pair, Bob's {qubits also collapse in exactly the same state as Alice's}. 
This would also occur if Alice measured in the Hadamard ($X$) basis instead. 
Then, if $N/2 + \delta N$ and $N/2 - \delta N$ are the number of qubits measured to be in states $\ket*{\uparrow}$ and $\ket*{\downarrow}$ respectively, {then each party ends up with} a finite ensemble of $N$ qubits with compressed density operator
\begin{equation} \label{eq:unbalanced_cdo}
    \varrho = \frac{\mathbb{I}}{2} + \frac{\delta N}{N} Z,
\end{equation}
where $\mathbb{I}$ is the identity operator in the single-qubit state space~\cite{Wang2005A}. 
In general, $\delta N$ is a random variable with a mean equal to zero and a standard deviation of $\sqrt{N}/2$. 
Therefore, in the limit $N\rightarrow \infty$, $\varrho$ converges to $\mathbb{I}/2${, 
since} for any $\epsilon,\delta>0$ there is a sufficiently large number $N_{\epsilon,\delta}$ such that $\Pr\left( \norm{\varrho-\mathbb{I}/2}_1 > \epsilon \right) < \delta$ for every $N>N_{\epsilon,\delta}$, where $\norm{\cdot}_1$ is the trace norm. 
This stochastic convergence is also expected when Alice measures in any other basis, and can be interpreted as the fact that two never-ending streams of independent random qubits in equiprobable orthogonal states cannot be physically distinguished~\cite{ZhanP14}.

The problem of the distinguishability of ensembles with the same compressed density operator (whether finite or infinite) has been addressed multiple times ever since the density operator formalism was introduced in detail by Fano in 1957~\cite{Fano1957,peres1993quantum,dEspagnat1995,PreskillNotes1998,Wang2005A,Long2006,ZhanP14,Bustamante2025}. 
Nowadays, density operators lie at the heart of quantum information theory, given their ability to represent both quantum and classical information present in physical states~\cite{Nielsen2010}. 
Despite this, many misconceptions regarding this formalism, mainly due to the wide (and often loose) usage of the world ``ensemble" in both quantum and statistical mechanics, have led to various interesting discussions and tutorials on the problem~\cite{Long2006,ZhanP14}. 
In one of such discussions, Preskill associated the indistinguishability of (finite) ensembles sharing the same (compressed) density operator with the impossibility of faster-than-light communication~\cite{PreskillNotes1998}. 
However, Wang {Chuan} \textit{et al.} pointed out an oversight in Preskill's analysis, which ultimately led them to propose the following QSDC protocol based on the distinguishability of finite ensembles with the same compressed density operator~\cite{Wang2005A,Wang2006}.

\subsection{The CDM06 protocol} \label{sec:wang_protocol}

The CDM06 protocol can be summarized in the following steps:
\begin{enumerate}[(i).]
    \item \textbf{Channel establishment:} Initially, Alice prepares an ensemble of $2n$ EPR pairs in state $\ket*{\phi^+_z}$ and sends one qubit from each pair to Bob. \label{step:channel_establishment}
    \item \textbf{Channel diagnosis:} Alice and Bob randomly choose $n$ pairs for security checking. 
    They both randomly measure their qubits in the $X$ and $Z$ bases and compare their outcomes for the rounds in which they measured in the same basis. 
    If there is full correspondence in their outcomes, then they can confirm the channel is safe and proceed. Otherwise they abort the protocol.
    \label{step:channel_diagnoses}
    \item \textbf{Entanglement distillation:} Bob chooses an entanglement distillation protocol and purify $n'$ out of the $n$ remaining pairs for both quantum error correction and privacy amplification.
    \label{step:entanglement_distillation}
    \item \textbf{Message establishment:} Alice encodes the bit of classical information she wants to send to Bob by performing a projective measurement on all of her qubits: if she wants to send a $0$ ($1$), she measures all of her qubits in the $Z$ ($X$) basis.
    \label{step:message_establishment}
    \item \textbf{Ensemble balancing:} After the measurement, both Alice and Bob end up with a random qubit string with an associated compressed density operator $\varrho$ of the form shown in Eq.~\eqref{eq:unbalanced_cdo}, using $X$ instead of $Z$ depending on Alice's choice of measurement basis. 
    Their goal during this step is for each party to get rid of the $\delta N$ excess qubits, thereby ``balancing" their ensembles. 
    To do this, Alice chooses and discards some of the qubits in her string so that the remaining string of $n''=2m$ qubits has exactly $m$ qubits in each eigenstate of the measured basis. 
    She tells Bob which qubits he must discard as well through a public authenticated one-way classical channel, so that each party end up with an ensemble of $2m$ qubits with compressed density operator $\mathbb{I}/2$.\label{step:ensemble_balancing}
    \item \textbf{Decoding:} Bob determines which bit was sent by Alice by measuring all of his qubits in the $Z$ basis. If he measures exactly $m$ qubits in states $\ket{\uparrow}$ and $\ket{\downarrow}$, then he infers Alice sent a $0$, otherwise he infers that Alice sent a $1$.\label{step:decoding}
\end{enumerate}
This protocol has two major disadvantages when compared to many QSDC protocols~\cite{Pan2024}. On the one hand, the protocol needs a large amount of quantum resources to securely transmit a single bit. This resourcefulness is quantified by the \textit{secure net bit rate} $R$ of the protocol, which is defined as the ratio of securely transmitted bits of useful information per elementary quantum resource (in this case, per EPR pair). On the other hand, even in the ideal case of 100\% efficient measurement devices, noiseless channels, a completely faithful source of EPR pairs and no eavesdropping, the protocol still has a non-vanishing \textit{average error probability} $P_e$ of the decoding process, namely
\begin{equation}
    P_e = \frac{1}{2^{2m+1}} \binom{2m}{m}
\end{equation}
for a given value of $m$~\cite{Wang2006}, whilst most known QSDC protocols, such as the standard DL04~\cite{Deng2004}, would have $P_e=0$. Some protocols are able to achieve this even with high-capacity transmission through superdense coding~\cite{Wang2005B}. Moreover, this error probability approaches zero only for increasing values of $m$, that is, at the cost of significantly lower values of $R$. Naturally, one would like to have a protocol with high and low values of $R$ and $P_e$, respectively, which makes the protocol overall less attractive for practical purposes.

The authors recognized these disadvantages in their article, concluding that the ``proposed protocol is more important in conceptual than practical''~\cite{Wang2006}. 
Here we extend their analysis by clarifying the main factors contributing to its limited resourcefulness, as well as by providing numerical values for secure net bit rates achievable by a generalized version of their protocol.

\subsection{Enhanced CDM06 resourcefulness} \label{sec:resourcefulness}

{In the following, we highlight opportunities to retain additional resources during the entanglement distillation and ensemble balancing steps.} 
On the one hand, canonical iterative entanglement distillation protocols~\cite{Bennett1996A,Deutsch1996} are generally known to require rather involved schemes and to be costly in terms of resources~\cite{Abdelkhalek2016}{ (although recent experimental advances have managed to evade some of these inconveniences and achieve significantly higher efficiencies}~\cite{Abdelkhalek2016,Ecker2021,Hu2021}).  
This cost is quantified by their yield or distillable entanglement $D_\infty(\rho)$, defined as the ratio between the amount of almost perfectly maximally entangled pairs that can be distilled from an asymptotically large amount of i.i.d. pairs in some mixed state $\rho$. 
{In the current context, if the mixed state $\rho$ of each EPR pair shared between Alice and Bob is diagonal in the Bell basis, then the yield $D_\infty(\rho)=\lim_{n\rightarrow\infty}n'/n$ achievable by a one-way entanglement distillation protocol is known to satisfy $D_\infty(\rho) \leq 1-S(\rho)$}, where $S$ is the von Neumann entropy function~\cite{Horodecki2004}. 
{This upper bound on $D_\infty(\rho)$ can be achieved asymptotically through the one-way hashing protocol}~\cite{Bennett1996B}{, although in realistic non-asymptotic scenarios,} actual ratios $n'/n$ are generally much lower~\cite{Fang2019}. 
On the other hand, the ensemble balancing step is an {innately} wasteful process, since {some of the} discarded qubits could instead be used to reduce $P_e$ or increase $R$. 
{It is worth noting that this step carries a $2/2^{n'}$ probability of requiring all $n'$ qubits to be discarded, which can compromise communication reliability.}
Thus, we conclude that a way in which the protocol can be made more resourceful (higher $R$) and correct (lower $P_e$) is by removing the entanglement distillation step and using all EPR pairs available for enhanced communication.

Omission of entanglement distillation, of course, comes with the cost of skipping quantum error correction and privacy amplification. 
However, if instead of aborting the protocol whenever an error is detected during channel diagnoses, Alice and Bob were allowed to choose a code for communicating after estimating the qubit error rates (QBERs) in the computational and Hadamard bases, then arbitrarily secure and correct direct communication would be possible {for low QBERs}, as we will show in Sec.~\ref{sec:security} in accordance to the quantum wiretap channel theory~\cite{Cai2004}. \
{Also, by removing entanglement distillation step, the receiving party is not required to perform unitary operations, making the protocol more suitable for real-life implementations of star network configurations}~\cite{Pan2024},{ where a central hub securely transmits data to multiple users with no capacity for coherent manipulation.}
That being said, we now proceed to propose a general model for one-way QSDC based on {the CDM06} protocol, where the local choice of measurement basis plays the role of the secret {bit}.

\section{Model for Quantum Secure Direct Communication}

{In this section we provide a state-based formulation of the proposed model for one-way QSDC with local choice of measurement basis as the secret bit. 
Then, based on this description, we certify information-theoretic security of four different {instances} of the model when the quantum channel is subject to BB84-symmetric attacks}~\cite{Ferenczi2012}{, in accordance to the quantum wiretap channel theory}~\cite{Cai2004}.

\subsection{State-based formulation} \label{sec:theoretical_model}

Let us now {provide a state-based mathematical formulation that completely encapsulates} {our proposed general model} for one-way QSDC using finite ensembles of $n$ shared EPR pairs per bit and a public authenticated classical channel, where the local choice of one of two mutually-unbiased measurement bases plays the role of the secret bit.
Choosing $n>1$ may seem not optimal at first, especially considering that many QSDC protocols use only one EPR pair per bit~\cite{Pan2024}. 
However, using more pairs per bit can enhance the protocol's secrecy, which may in turn increase its overall secure net bit rate $R$, as will be shown to happen later for the case $n=2$.

In the proposed model, both the encoding and decoding of classical information in quantum systems are implemented by measurements in either the computational or the Hadamard basis. 
Also, it is presupposed that the two communicating parties share an insecure one-way quantum channel, wiretapped by an eavesdropper who interacts independently and identically with all the transmitted qubits, and who is able to measure their probe quantum systems collectively per finite ensemble in a source-replacement scheme~\cite{Ferenczi2012}. 
This model will generalize the CDM06 protocol when the entanglement distillation process is removed by allowing the parties to use the public authenticated classical channel in any arbitrary way they choose (rather than just to signal the excess qubits as in step~\ref{step:ensemble_balancing}), and to choose an arbitrary decoding measurement (instead of one constrained by step~\ref{step:decoding}). 
As such, there is no need to wastefully discard resources in our model.

It is important to note here that a protocol for one-way QSDC using the choice of measurement basis as the secret has been previously proposed by Liu \textit{et al.}~\cite{Liu2014}, based on a previous work by Kalev \textit{et al}.~\cite{Kalev2013}. 
However, their security proof focuses on the zero-error case, providing a foundation that is yet to be extended to more realistic scenarios.
Furthermore, their protocol requires a two-way quantum channel (despite the fact that information is only being transmitted in one direction), whilst our model requires only a one-way quantum channel, making it particularly suitable for QSDC in star network configurations~\cite{Pan2024}.

The model can be formulated as follows: the sender (Alice) prepares a sufficiently large amount of EPR pairs in the maximally entangled state $\ket{\phi^+_z}$. Let $A_i$ and $B_i$ represent the qubits composing the $i$-th EPR pair prepared by Alice. 
{She} then sends each $B_i$ qubit to the receiver (Bob) through an insecure quantum channel, where they are intercepted by an eavesdropper (Eve) who repeatedly performs some fixed unitary $U$ between each $B_i$ qubit and different identical probe systems $E_i$ of hers. 
Eve then resends the $B_i$ qubits to Bob, such that Alice and Bob end up sharing many copies of a mixed state $\rho_{\mathfrak{ab}} = \Tr_{\mathfrak{e}}(\ketbra*{\psi}_{\mathfrak{abe}})$, where
\begin{equation}\label{eq:generic_purification}
    \ket*{\psi}_{\mathfrak{abe}} =(\mathbb{I}_{\mathfrak{a}} \otimes U_{\mathfrak{be}}) \ket*{\phi^+_z}_{\mathfrak{ab}} \ket{e}_{\mathfrak{e}},
\end{equation}
for any subsystem $\mathfrak{abe}=A_iB_iE_i$ and some arbitrary initial state {of Eve's probe} $\ket{e}$. 
Bob announces that he has received the qubits, then he and Alice proceed to use some of these resources to estimate the QBERs $Q_Z$ and $Q_X$ in the $Z$ and $X$ bases, respectively, as in standard BB84 parameter estimation~\cite{Ramona2021}. 
This will allow them to diagnose their quantum channel and, consequently, either abort the session if the QBERs are too high, or choose an arbitrarily secure and correct code to proceed with communication, as will be explained later in Sec.~\ref{sec:security}. 
Since no entanglement distillation protocol is performed, the remaining pairs {will be} used for communication. 
{What we propose is for these pairs to} be divided into ensembles of $n$ pairs, each of which will be used for transmitting a single bit of classical information. 
In other words, Alice and Bob will use $n$ {EPR pairs per classical bit transmitted. These bits are not necessarily bits of explicit useful information (also called \textit{net bits}), but may be determined by a particular codebook which Alice and Bob are free to agree on to securely and correctly communicate after characterizing the channel.}
If $C$ is the rate of their codebook, then $R=C/n$ {secure net bits} per EPR pair are being transmitted on average.

\begin{figure}[h]
    \centering
    \includegraphics[width=\linewidth]{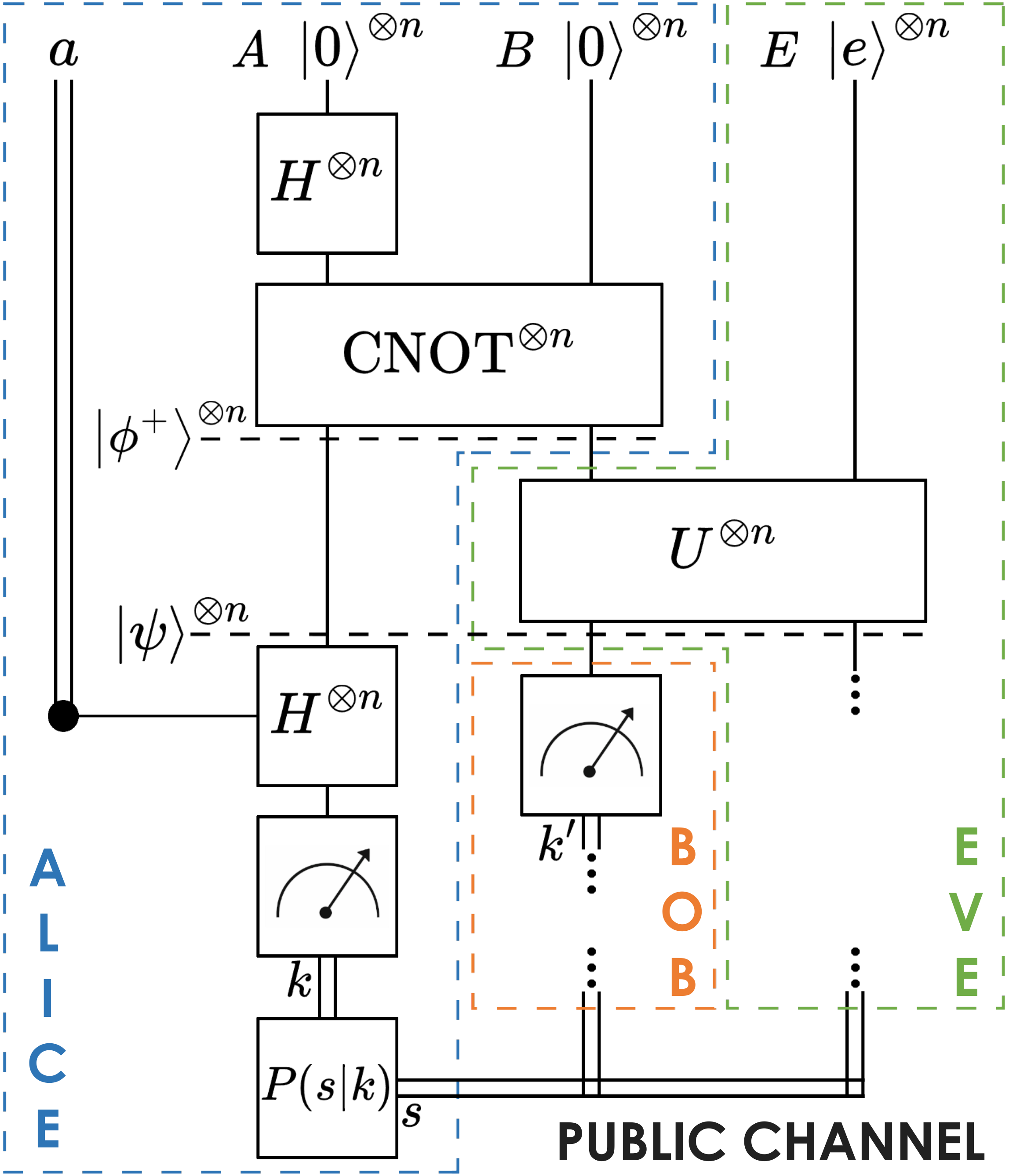}
    \caption{{Quantum circuit for transmission of a classical bit $a$. Alice prepares $n$ EPR pairs $\ket*{\phi^+}$ and sends one qubit from each pair to Bob through an insecure quantum channel. Eve intercepts each qubit, repeatedly performing a unitary $U$ between them and different identical probe systems of hers, then forwards them to Bob. Bob measures all received qubits in the computational basis, obtaining outcome $k'$. Alice measures her qubits in the computational ($a=0$) or the Hadamard ($a=1$) basis, obtaining outcome $k$, then publicly announces a bit string $s$ following a predefined disclosure strategy $P(s|k)$. Colored dashed boundaries indicate each party's lab, and ellipses ($\dots$) are used to indicate that further processing of incoming information by a party is left unspecified.}}
    \label{fig:qcircuit}
\end{figure}

Now, to send a classical bit $a=0,1$ using one of these finite ensembles, Alice measures all of her $n$ qubits in either the $Z$ (for $a=0$) or the $X$ (for $a=1$) basis, which collapses her $n$-qubit system into a state $\ket*{k^{(a)}}$, where $k=k_1\dots k_n$ is a random bit string denoting the outcome of her measurement, following the notation
\begin{equation*}
    \ket*{0^{(a)}} = \begin{cases}
        \ket*{\uparrow} & a=0 \\
        \ket*{\rightarrow} & a=1 
    \end{cases} \qquad
    \ket*{1^{(a)}} = \begin{cases}
        \ket*{\downarrow} & a=0 \\
        \ket*{\leftarrow} & a=1 
    \end{cases},
\end{equation*}
for instance, $\ket*{010^{(0)}}=\ket{\uparrow \downarrow \uparrow}$, while $\ket*{010^{(1)}}=\ket{\rightarrow \leftarrow \rightarrow}$. 
At the moment, Bob has no way to infer Alice's bit without some additional information. 
Therefore, based solely on her measurement outcome $k$, she randomly announces a bit string $s$ according to an agreed conditional distribution $P(s|k)$, through a public authenticated classical channel $\mathcal{C}$. 
If $P_\mathcal{A}(a)$ is the probability of Alice transmitting a classical bit $a$, then the state of the whole system at this point is
\begin{equation} \label{eq:big_state}
\begin{gathered}
    \sum_a P_\mathcal{A}(a) \ketbra*{a}_\mathcal{A} \otimes \frac{1}{2^n} \sum_{k} \ketbra*{k^{(a)}}_A \\ \otimes \sum_s P(s|k) \ketbra*{s}_\mathcal{C} \otimes \ketbra*{\psi}_{BE|a,k}.
\end{gathered}
\end{equation}
{Here, the first two terms in the tensor product represent the states of Alice's classical register $\mathcal{A}$ and quantum system $A=A_{i_0}\dots A_{i_{n-1}}$, the third term represents the information disclosed through the public channel $\mathcal{C}$ and }
\begin{equation}
    \ket{\psi}_{BE|a,k} = U^{\otimes n} \ket*{k^{(a)}}_B \ket{e}^{\otimes n}_E
\end{equation}
{is the state of Bob and Eve's quantum systems $B=B_{i_0}\dots B_{i_{n-1}}$ and $E=E_{i_0}\dots E_{i_{n-1}}$ right after Alice's measurement.} 
We further assume that Bob performs {either a computational or a Hadamard basis measurement} on his qubits, and thus the resulting state of the whole system becomes
\begin{equation} \label{eq:final_state}
\begin{gathered}
    \rho_{\mathcal{A}A\mathcal{C}BE} = \sum_a P_\mathcal{A}(a) \ketbra*{a}_\mathcal{A} \otimes \frac{1}{2^n}\sum_{k}  \ketbra*{k^{(a)}}_{A} \\ \otimes \sum_{s} P(s|k) \ketbra*{s}_\mathcal{C} 
     \otimes \sum_{k'}  P_{a}(k'|k) \ketbra*{k'^{(b)}}_B \\ \otimes \ketbra*{\psi}_{E|a,k, k'}, 
\end{gathered}
\end{equation}
where $b$ represents Bob's fixed measurement basis, $\ket{\psi}_{E|a,k, k'}$ is the conditional post-measurement state of Eve's quantum system $E$, and $P_{a}(k'|k)$ is the probability of Bob's measurement outcome $k'$ conditioned on Alice's measurement outcome $k$ and choice of basis $a$. 
When Alice and Bob measure in the same basis (i.e. when $a=b$), this conditional probability can be written in terms of the QBERs as
\begin{equation} \label{eq:bob_outcome_stats}
    P_a(k'|k)=
        Q_a^{|k\oplus k'|}(1-Q_a)^{n-|k\oplus k'|},
\end{equation}
where $|\cdot|$ is the Hamming weight function and we have used notation $Q_0=Q_Z$ and $Q_1=Q_X$ for convenience. 
{A quantum circuit diagram for the transmission of a single classical bit $a$ using the proposed model with $b=0$ is shown in Fig.}~\ref{fig:qcircuit}. 

We note that the state $\rho_{\mathcal{A}A\mathcal{C}BE}$ in Eq.~\eqref{eq:final_state} {entirely encapsulates our proposed model. 
However, to be able to study and certify its security} from an information-theoretic standpoint, we must first specify how Alice and Bob choose to use the public channel. 
Let $\mathcal{K}(s)$ be the set of all bit strings $k$ that are compatible with a given $s$, that is, the set of all $k$ for which $P(s|k) \neq 0$. 
Conversely, let $\mathcal{S}(k)$ be the set of all $s$ for which $P(s|k) \neq 0$. 
In this work, we consider {four instances of the model given by the following disclosure strategies}: 
\begin{itemize}
    \item \textbf{Full Outcome Disclosure:} Alice publicly announces her outcome $k$ through $\mathcal{C}$, i.e., $s=k$. This is the case where the maximum amount of information is revealed through the public channel. 
    Here, for a fixed $s$ and any $k\in\mathcal{K}(s)$, one has $|\mathcal{K}(s)|=|\mathcal{S}(k)| = 1$, which is evident from the fact that $P(s|k)=1$ if and only if $s=k$.
    \item \textbf{Excess Bits Disclosure:} Let $m_1$ and $m_0$ be the number of ones and zeroes in $k$. 
    Then, Alice randomly chooses $|m_1-m_0|$ of the excess bits that unbalance $k$ and announces their position. 
    Thus, we define $s$ to be an $n$-bit string where every $i=1,\dots,n$ for which $s_i=1$ indicates the position of a chosen excess qubits. 
    Here, for a fixed $s$ and any $k\in\mathcal{K}(s)$ one has
    \begin{align*}
        |\mathcal{K}(s)|&=\binom{n-|s|}{(n-|s|)/2}f_s \\ |\mathcal{S}(k)| &= \binom{(n+|s|)/2}{|s|},
    \end{align*}
    where $f_s$ is a factor equal to $1$ if $s$ is an all-zero bit string and $2$ otherwise. 
    Noticeably, this case corresponds to {the CDM06} protocol when Bob chooses to discard the excess qubits and decode according to step~\ref{step:decoding} in Sec.~\ref{sec:wang_protocol}.
    \item \textbf{Weight Disclosure:} Alice publicly announces the number of ones in $k$, that is, $s=|k|$. {In contrast to previous strategies}, where $s$ is a bit string of length $n$ indicating the position of certain key bits in $k$, $s$ is now an integer number indicating a global property of $k$. Here, for a fixed $s$ and any $k\in\mathcal{K}(s)$ one has $|\mathcal{K}(s)|=\binom{n}{s}$ and $|\mathcal{S}(k)|=1$.
    \item \textbf{Parity Disclosure:} Alice publicly announces the parity of $k$, i.e., $s=0$ ($1$) if $k$ has an even (odd) number of ones. Here, for a fixed $s$ and any $k\in\mathcal{K}(s)$ one has $$|\mathcal{K}(s)|=\sum_{l=0}^{\lfloor n/2 \rfloor}\binom{n}{2l+s}$$ and $|\mathcal{S}(k)|=1$.
\end{itemize}
All these alternatives satisfy two useful properties. First, for a given measurement outcome $k$, all compatible public announcements are equiprobable, i.e. $P(s|k)=1/|\mathcal{S}(k)|$ for all $s\in\mathcal{S}(k)$. Second, that for a fixed public announcement $s$, the cardinality $|\mathcal{S}(k)|$ is equal for all compatible measurement outcomes $k\in\mathcal{K}(s)$. These two properties allows us to compute the posterior probabilities $P(s)$ and $P(k_s|s)$ for a fixed $s$ and any $k_s\in\mathcal{K}(s)$ as
\begin{gather}
    P(s)=\frac{1}{2^n}\sum_k P(s|k)=\frac{|\mathcal{K}(s)|}{2^n |\mathcal{S}(k_s)|} \\
    P(k_s|s) = \frac{P(s|k_s)}{\sum_k P(s|k) }= \frac{1}{|\mathcal{K}(s)|},
\end{gather}
which will be used to evaluate the security of the model in Sec.~\ref{sec:security}. Evidently, many other different uses of the public channel can be proposed and the two discussed properties do not need to be satisfied.

\subsection{Information-theoretic security under\\BB84-symmetric attacks} \label{sec:security}
Information-theoretic security of our QSDC model, when viewed as a classical-quantum wiretap channel, can be ensured by proving the strict positivity of its \textit{secrecy capacity} $C_s$, since this would imply the existence of arbitrarily secure and correct codes, with code rates up to the secrecy capacity~\cite{Cai2004}. Thus, a positive $C_s$ would mean that, after characterizing the channel, Alice and Bob can choose a code (that is, a codebook and a decoding measurement, as defined in reference~\cite{Cai2004}) with rate $C < C_s$ and use it to transmit $R=C/n$ secure bits of useful information per EPR pair. Since each ensemble is composed by $n$ EPR pairs, then $C$ can also be understood as the number of secure bits of useful information that can be transmitted per ensemble. Theorem 4 in reference~\cite{Cai2004} states that the secrecy capacity of a classical-quantum wiretap channel has a lower bound given by
\begin{equation} \label{eq:secrecy_capacity_bound}
    C_s \geq \max_{P_\mathcal{A}} ( \chi_B-\chi_E ),
\end{equation}
where $\chi_B$ and $\chi_E$ are the Holevo quantities associated to Bob and Eve respectively, which in this case are equal to the quantum mutual informations $I(\mathcal{A};\mathcal{C}B)$ and $I(\mathcal{A};\mathcal{C}E)$ determined by the final state in Eq.~\eqref{eq:final_state}. If this lower bound is strictly greater than zero, it is then equal to the rate $C$ of some code that Alice and Bob can use for secure direct communication through our model. 

Ideally, we would like to consider the case where the eavesdropper performs an optimal attack, that is, a unitary $U$ that maximizes Eve's accessible information of Alice's register while being compatible with QBERs $Q_Z$ and $Q_X$. However, finding such an optimal attack is not trivial. 
Thus, we choose to study the security of our QSDC model under a specific class of attacks, which we call BB84-symmetric in accordance to reference~\cite{Ferenczi2012}. 
We define these attacks as those that are compatible with fixed QBERs in the $Z$ and $X$ bases, and optimal for quantum key distribution protocols that use the same signal states as {the proposed model}, i.e. states $\ket*{k_i^{(a)}}=\ket{\uparrow},\ket{\downarrow},\ket{\rightarrow},\ket{\leftarrow}$, just as in BB84. 
These attacks are known to leave the reduced state $\rho_{\mathfrak{ab}}$ diagonal in the Bell basis~\cite{Ferenczi2012}. 
Were we to consider a variation of {the model}, where Alice and Bob perform a discrete twirling operation~\cite{Bennett1996B}, then $\rho_{\mathfrak{ab}}$ would be diagonal in the Bell basis regardless of Eve's attack, in which case we could certify information-theoretic security in the worst-case scenario. 
However, we do not consider this variation in order to avoid extra unitary operations to be performed by the receiving party, so that the protocol can be easily implemented in real-life star network configurations~\cite{Pan2024}.

Thus, assuming $\rho_{\mathfrak{ab}}$ to be diagonal in the Bell basis, we use a purification with Schmidt form
\begin{equation}
    \ket{\psi}_{\mathfrak{abe}} = \sum_{i,j=0}^1 \sqrt{\lambda_{ij}}\ket{\Phi_{ij}}_{\mathfrak{ab}} \ket{e_{ij}}_\mathfrak{e},
\end{equation}
where the $\ket{\Phi_{ij}}=(X^i Z^j \otimes \mathbb{I})\ket{\phi^+_z}$ states run over all Bell states, and the squared Schmidt coefficients $\lambda_{ij}$ represent a probability distribution satisfying $Q_Z=\lambda_{10} + \lambda_{11}$ and $Q_X = \lambda_{01} + \lambda_{11}$. After some algebra, one finds that $\chi_B$ can be computed in terms of the conditional probability distribution $P_a(k'|s) = \sum_k P_a(k'|k)P(k|s)$ as
\begin{equation}
\begin{split}
    \chi_B = \sum_s P(s) \Bigg[ H \bigg(\bigg\{\sum_a P_\mathcal{A}(a)P_a(k'|s)\bigg\}_{k'}\bigg)  \\ - \sum_a P_\mathcal{A}(a) H(\{P_a(k'|s)\}_{k'})\Bigg],
\end{split}
\end{equation}
where now $P_a(k'|k)=1/2^n$ whenever $a\neq b $.
On the other hand, $\chi_E$ is given by
\begin{equation}
\begin{split}
    \chi_E = \sum_s P(s) \Bigg[ S\bigg(\sum_a P_\mathcal{A}(a) \rho_{E|a,s}\bigg) \\ - \sum_a P_\mathcal{A}(a) S(\rho_{E|a,s}) \Bigg],
\end{split}
\end{equation}
with the conditional state $\rho_{E|a,s}$ given by
\begin{equation}
\begin{split}
    \rho_{E|a,s} = \sum_{k,r} P(k|s) \bigotimes_{i=1}^n \ketbra*{\xi_{r_i}^{(a)}(k_i)}_\mathfrak{e},
\end{split}
\end{equation}
where both $k$ and $r$ run over all bit strings of length $n$ and
\begin{subequations}
\begin{align}
    & \ket*{\xi_{r_i}^{(0)}(k_i)} = \sqrt{\lambda_{r_i 0}}\ket*{e_{r_i 0}} + (-1)^{k_i \oplus r_i}\sqrt{\lambda_{r_i 1}}\ket*{e_{r_i 1}} \\
    &\ket*{\xi_{r_i}^{(1)}(k_i)} = \sqrt{\lambda_{0 r_i}}\ket*{e_{0r_i}} + (-1)^{k_i}\sqrt{\lambda_{1 r_i}}\ket*{e_{1 r_i}}
\end{align}
\end{subequations}
are sub normalized states satisfying $\braket*{\xi_{r_i}^{(a)}(k_i)}{\xi_{r'_i}^{(a)}(k_i)} = 0$ for $r_i\neq r_i'$. For fixed $Q_Z$ and $Q_X$, the squared Schmidt coefficients $\lambda_{ij}$ admit a parametrization in terms of a single real variable $t$ controlled by Eve~\cite{Glaucia2025}, namely
\begin{subequations}
    \begin{align*}
        &\lambda_{00}=1-\frac{Q_X+t+Q_Z}{2} & & \lambda_{01}= \frac{Q_X+t-Q_Z}{2} \\
        &\lambda_{10}= \frac{-Q_X+t+Q_Z}{2} & & \lambda_{11}= \frac{Q_X-t+Q_Z}{2}.
    \end{align*}
\end{subequations}
\begin{figure}[t]
    \centering
    \includegraphics[width=1\linewidth]{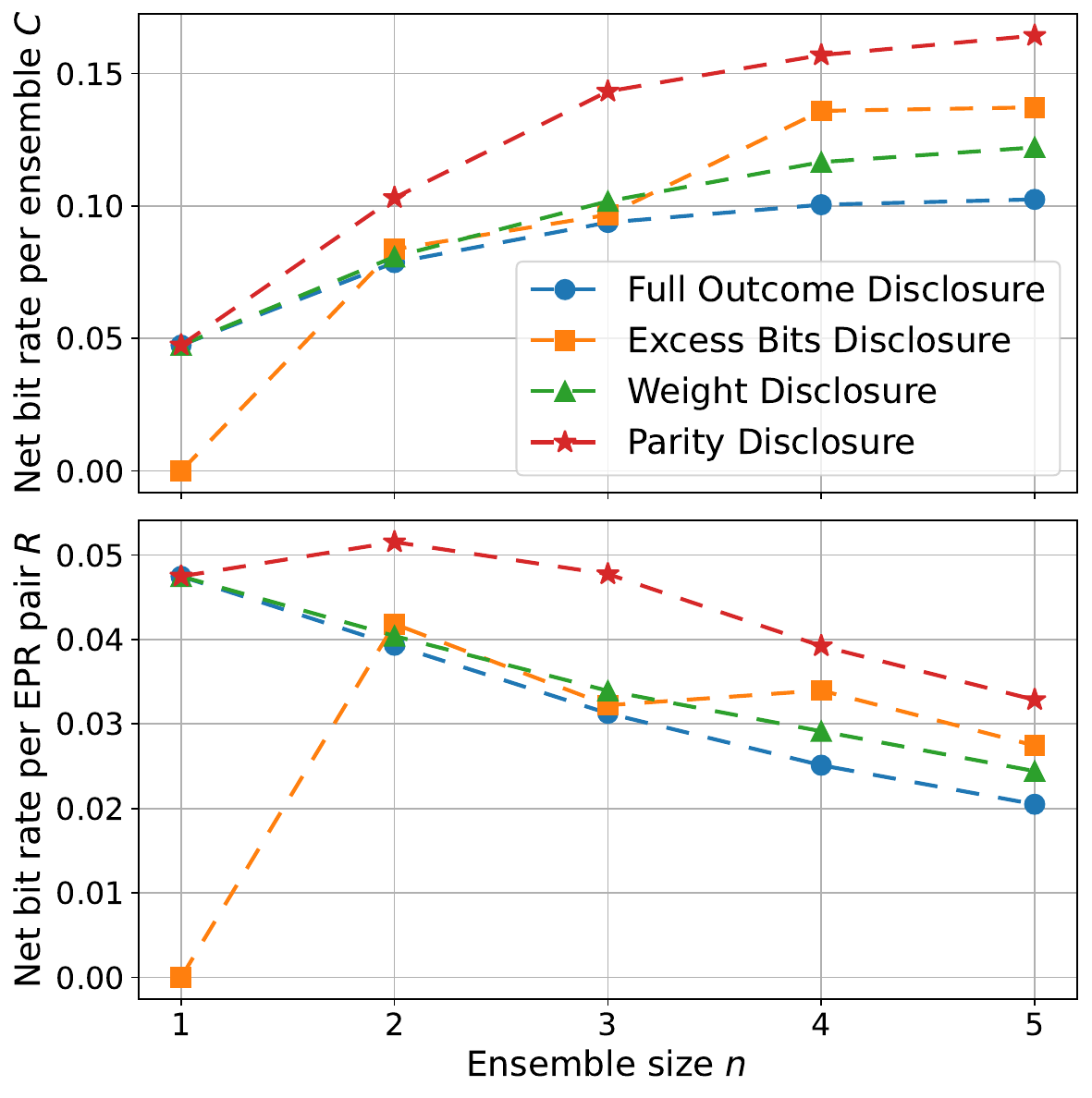}
    \caption{Secure net bit rates per ensemble (top) and per EPR pair (bottom) achievable by four different {instances of the model} as functions of the size of the ensembles. These rates were computed for a one-way quantum channel with fixed qubit error rates $Q_Z=Q_X=0.05$.}
    \label{fig:SNBR_vs_n}
\end{figure}
Therefore, to certify information-theoretic security of {the model} when the quantum channel is subject to optimal BB84-symmetric attacks, we use the Nelder-Mead numerical optimization method from the SciPy Python library to compute achievable code rates $C$ given by the lower bound in Eq.~\eqref{eq:secrecy_capacity_bound}, using a value of $t$ that minimizes the difference $\chi_B-\chi_E$ before maximizing in $P_{\mathcal{A}}$. Then, whenever we find an strictly positive $C$, we ensure the existence of an arbitrarily secure and correct code for communication, despite not providing its construction. 
Achievable secure net bit rates per ensemble ($C$) and per EPR pair ($R=C/n$) for the four {disclosure strategies} as functions of the ensemble size $n$ for fixed QBERs $Q_Z=Q_X=0.05$ are shown in Fig.~\ref{fig:SNBR_vs_n}. 
In this figure we can see an expected trade-off between the achievable rates and the ensemble size: as the ensemble size $n$ grows, arbitrarily secure and correct codes with higher code rates $C$ are shown to exist, however, their secure net bit rates per EPR pair $R=C/n$ can end up decreasing if $C$ grows slower than $n$, which happens to all four alternatives for $n\geq 2$. In fact, all four achieve their maximum rate $R$ for $n=2$, the highest among all of them being $R\approx0.052$ corresponding to the Parity Disclosure {strategy}. For the same fixed QBERs, the standard and modified DL04 protocols are able to achieve rates up to $R\approx0.245$ and $R\approx0.427$ respectively, that is, around 5 and 8 times higher than the best one found for our model~\cite{Wu2019}. 
This means that using finite ensembles of shared EPR pairs per bit does not necessarily allow for enhanced secure net bit rates, as expected from a similar conclusion made by Wang Chuan \textit{et al.} about the CDM06 protocol upon which {the proposed model} is based.

\begin{figure}[t]
    \centering
    \includegraphics[width=1\linewidth]{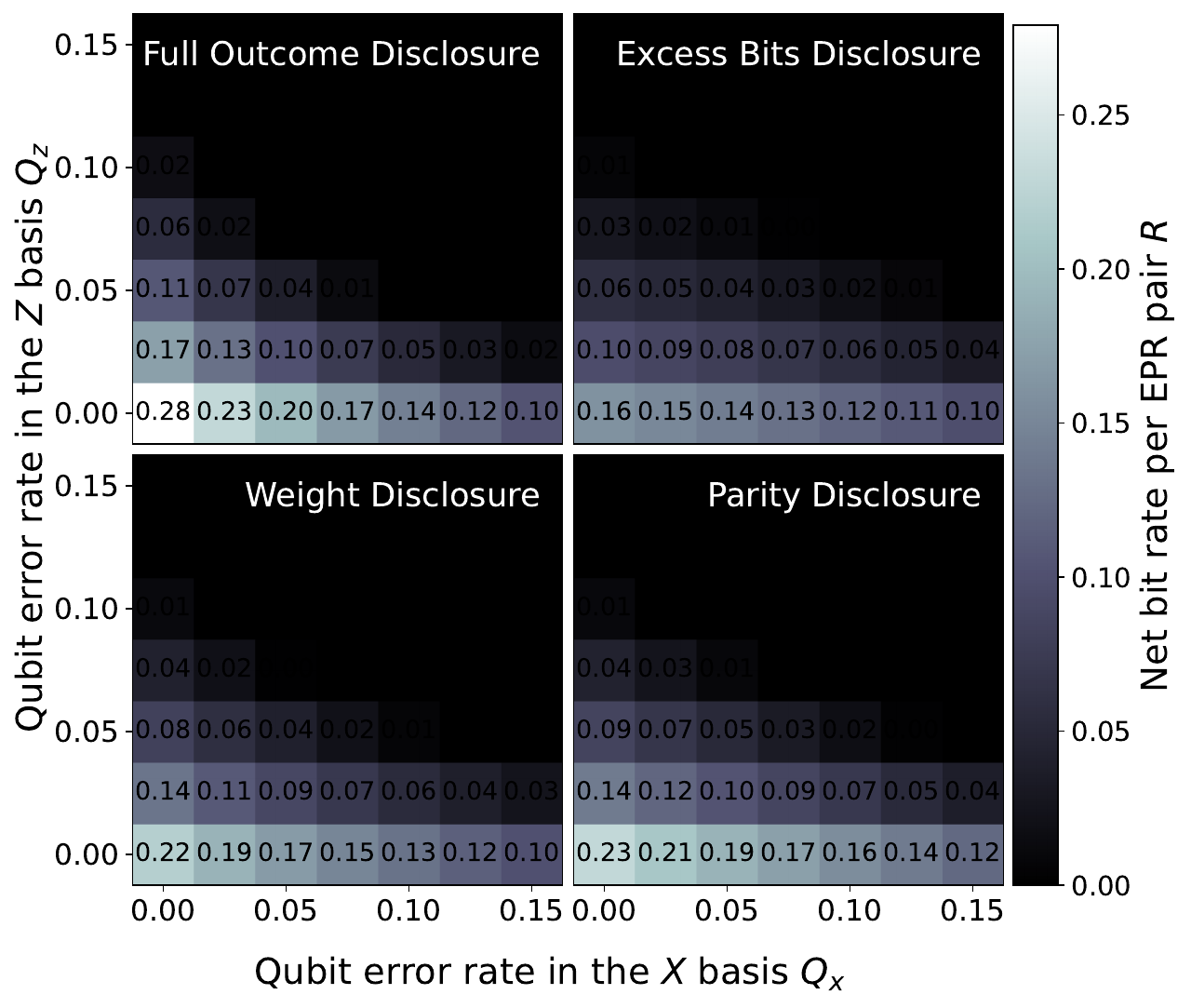}
    \caption{Colormaps representing the secure net bit rates per EPR pair achievable by four different {instances} of {the proposed model} as functions of the qubit error rates in the $Z$ and $X$ bases. The colormaps suggest that the model is more robust to phase-flip than to bit-flip errors, as expected from its asymmetry. For sufficiently high qubit error rates (corresponding to the completely dark regions), secure communication cannot be ensured. All rates were computed for an ensemble size of $n=2$.}
    \label{fig:Bound_Map}
\end{figure}
In Fig.~\ref{fig:Bound_Map}, four colormaps representing the achievable secure net bit rates per EPR pair $R$ of the four {instances of the model} as functions of the QBERs in the $Z$ and $X$ bases for a fixed ensemble size of $n=2$ are shown. Here we have fixed $b=0$, meaning Bob always measures in the computational basis. If the QBERs in the channel are found to correspond to a completely dark region (where strict positivity of the secrecy capacity is not ensured), then the parties must abort communication, given that in such cases information-theoretic security cannot be certified. The maximum rates found are virtually always achieved by either the Full Outcome Disclosure or the Parity Disclosure {strategies}. In a zero-error scenario, the maximum rate $R\approx0.279$ {was achieved using the Full Outcome Disclosure strategy}, which is still less than a third of the $R=1$ rate achievable by the standard DL04 protocol. More importantly, the colormaps suggest that {the proposed model} is more robust to phase-flip than to bit-flip errors (which correspond to QBERs in the $X$ and $Z$ basis respectively), given that the achievable rate $R$ decreases slower for increasing $Q_X$ than for increasing $Q_Z$. This is expected from the asymmetry of the model: since Bob only measures in the $Z$ basis, phase-flip errors do not affect his measurement outcome statistics, as reflected by the conditional probability distribution $P_a(k'|k)$. Had we instead fixed $b=1$, the model would have shown to be more robust to bit-flip than to phase-flip errors in a completely analogue fashion.

\section{Conclusions} \label{sec:conclusions}

We have demonstrated a model for one-way QSDC using finite ensembles of shared EPR pairs per bit and a public authenticated classical channel, in which the local choice of measurement basis plays the role of the secret. 
Within the framework provided by the quantum wiretap channel theory, this model was shown to allow for arbitrarily secure and correct communication when the quantum channel is subject to BB84-symmetric attacks for sufficiently low QBERs.
Were it proven that BB84-symmetric attacks are optimal in terms of the eavesdropper accessible information about the secret, then information-theoretic security of {the proposed model} would be ensured even in the worst-case scenario. 
The asymmetry of the model makes it robust against either phase-flip or bit-flip errors, as long as the receiver measures in the computational or Hadamard basis, respectively. 
The best achievable secure net bit rates per EPR pair given by the four {disclosure strategies considered in this work} were, at best, between four and five times lower than the rates given by the standard DL04 protocol for QSDC. 
This leads us to conclude that using finite ensembles of shared EPR pairs per bit does not necessarily allow for enhanced secure net bit rates, as expected from the work of Wang Chuang \textit{et al.} upon which this research is based~\cite{Wang2006}. 
However, unlike most models for one-way QSDC in the literature~\cite{Pan2024}, our model requires neither a two-way quantum channel nor coherent manipulation to be done by either party, only computational and Hadamard basis measurements. 
Moreover, since the receiver is only required to perform computational basis measurements, {the proposed model} is particularly suitable for real-life implementations of secure direct communication in star network configurations. 
Many different {instances} and variations of {the model}, which may yield higher secure net bit rates, are yet to be explored.

\section*{Acknowledgments}

We thank Prof. Gl\'aucia Murta, whose knowledge and insights in quantum cryptography helped us shape this work during S.B. internship at the Quantum Optics and Quantum Information Research Unit of the Atominstitut. 
S.B. gratefully acknowledges funding from the project “Ampliación del uso de la mecánica cuántica desde el punto de vista experimental y su relación con la teoría, generando desarrollos en tecnologías cuánticas útiles para metrología y computación cuántica a nivel nacional”, BPIN 2022000100133, from Sistema General de Regalías (SGR) of Minciencias, Gobierno de Colombia.
E.A.\ acknowledges financial support from the  the Austrian Research Promotion Agency (FFG) through the Project NSPT-QKD FO999915265.

\bibliography{references}

\end{document}